\documentclass[sigconf,nonacm]{aamas}

\usepackage{balance} 

\makeatletter
\gdef\@copyrightpermission{
  \begin{minipage}{0.2\columnwidth}
   \href{https://creativecommons.org/licenses/by/4.0/}{\includegraphics[width=0.90\textwidth]{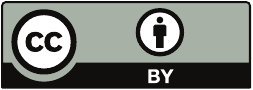}}
  \end{minipage}\hfill
  \begin{minipage}{0.8\columnwidth}
   \href{https://creativecommons.org/licenses/by/4.0/}{This work is licensed under a Creative Commons Attribution International 4.0 License.}
  \end{minipage}
  \vspace{5pt}
}
\makeatother

\setcopyright{ifaamas}
\acmConference[Preprint]{Preprint}{under review}{October 2026}{}
\copyrightyear{2027}
\acmYear{2027}
\acmDOI{}
\acmISBN{}

\usepackage{booktabs}
\usepackage{graphicx}
\usepackage{xcolor}
\usepackage{pgfplots}
\pgfplotsset{compat=1.18}
\definecolor{cbest}{HTML}{2a78d6}     
\definecolor{ckeep}{HTML}{eb6834}     
\definecolor{cforget}{HTML}{1baf7a}   
\definecolor{cwhole}{HTML}{eda100}    
\definecolor{cassist}{HTML}{e87ba4}   
\definecolor{cpromise}{HTML}{008300}  
\definecolor{cdir}{HTML}{4a3aa7}      
\definecolor{cgrey}{HTML}{52514e}     
\definecolor{cnewest}{HTML}{8d6e3f}   
\definecolor{cclosest}{HTML}{1aa3b8}  
\pgfplotsset{
  paperaxis/.style={height=4.1cm, grid=major, grid style={gray!18, very thin},
    axis line style={gray!55}, tick label style={font=\scriptsize, color=black!75},
    label style={font=\scriptsize}, title style={font=\scriptsize\bfseries, yshift=-0.6ex},
    every axis plot/.append style={line width=1.1pt, mark size=1.9pt},
    legend style={font=\scriptsize, draw=none, fill=none, /tikz/every even column/.append style={column sep=0.5em}},
    clip=false, xlabel near ticks, ylabel near ticks},
  sbest/.style={color=cbest, mark=*},
  skeep/.style={color=ckeep, mark=square*},
  sforget/.style={color=cforget, mark=triangle*},
  swhole/.style={color=cwhole, mark=diamond*},
  sassist/.style={color=cassist, mark=pentagon*},
  snone/.style={color=cgrey, densely dashed, mark=x, mark options={solid}},
  spromise/.style={color=cpromise, mark=otimes*},
  sdir/.style={color=cdir, mark=*},
  swide/.style={color=cgrey, mark=triangle*},
  snewest/.style={color=cnewest, mark=square*},
  sclosest/.style={color=cclosest, mark=diamond*},
}
\tikzset{endlab/.style={font=\tiny, color=black!75, anchor=west, xshift=2pt, inner sep=1pt}}
\usetikzlibrary{positioning,arrows.meta,calc,fit}
\usepgfplotslibrary{groupplots}

\definecolor{tkhead}{HTML}{B65C27}   
\definecolor{tkbody}{HTML}{FCF4EC}   
\definecolor{tktitle}{HTML}{FFFFFF}  
\definecolor{cchead}{HTML}{E3EDE5}   
\definecolor{ccbody}{HTML}{F7FAF7}   
\definecolor{ccframe}{HTML}{B4CBBA}  
\definecolor{cctitle}{HTML}{2E5940}  
\pgfdeclarelayer{boxback}\pgfsetlayers{boxback,main}
\newcommand{\resultbox}[7]{\par\addvspace{\medskipamount}\noindent
  \begin{tikzpicture}
    \node[anchor=north west, text width=\dimexpr\columnwidth-14pt\relax,
      inner xsep=6pt, inner ysep=3.5pt, font=\small\bfseries, text=#5]
      (rbt) at (0,0) {#6};
    \node[anchor=north west, text width=\dimexpr\columnwidth-14pt\relax,
      inner xsep=6pt, inner ysep=4.5pt, font=\small]
      (rbb) at (rbt.south west) {#7};
    \begin{pgfonlayer}{boxback}
      \begin{scope}
        \clip[rounded corners=3pt] (rbt.north west) rectangle (rbb.south east);
        \fill[#2] (rbt.north west) rectangle (rbb.south east);
        \fill[#1] (rbt.north west) rectangle (rbt.south east);
      \end{scope}
      \draw[#3, line width=#4, rounded corners=3pt]
        (rbt.north west) rectangle (rbb.south east);
    \end{pgfonlayer}
  \end{tikzpicture}\par\addvspace{\medskipamount}}
\newcommand{\takeaway}[2]{\resultbox{tkhead}{tkbody}{tkhead}{0.8pt}{tktitle}{#1}{#2}}
\newcommand{\conclusion}[2]{\resultbox{cchead}{ccbody}{ccframe}{0.6pt}{cctitle}{#1}{#2}}

\AtBeginDocument{%
  }
\graphicspath{{figures/}}

\begin{document}

\title{The Inbox as a Coordination Primitive
for LLM Agent Societies}

\author{Karthik Sreedhar}
\email{ks4190@columbia.edu}
\affiliation{%
  \institution{MIT Media Lab}
  \city{Cambridge}
  \country{USA}}

\author{Leo Yuan}
\affiliation{%
  \institution{MIT Media Lab}
  \city{Cambridge}
  \country{USA}}

\author{Pradyumna Chari}
\affiliation{%
  \institution{MIT Media Lab}
  \city{Cambridge}
  \country{USA}}

\author{Ramesh Raskar}
\affiliation{%
  \institution{MIT Media Lab}
  \city{Cambridge}
  \country{USA}}

\begin{abstract}
Language-model agents increasingly coordinate in large groups, and current agent protocols focus on delivering messages. For these agents, however, the scarce resource is not delivery but reading: every message costs tokens, and the context is finite. Current infrastructure works by push: the sender decides when a message enters the recipient’s context, so the recipient reads in arrival order and drops the oldest messages once its context is full. We study the alternative, pull, in which messages wait in an inbox outside the context and the recipient’s own rules decide what it reads and keeps. We argue that the inbox is a coordination primitive: each agent sets its rules privately, and those rules decide whether other agents’ requests reach it. In simulated societies of up to 16,000 agents, with every decision an LLM judgment, we report three findings. First, push fails as an agent takes part in more coordination at once: an agent leading 20 negotiations with a full memory proposes options already rejected 97\% of the time, because it has dropped the oldest replies. Second, an inbox is necessary under heavy, noisy traffic. Where first come, first read misses every request, an inbox that first reads messages carrying a refundable deposit still finds about 60\% of the agents a request needs, at every size from 1,000 to 16,000. On a real email network, losses compound: when half the agents adopt such an inbox, forwarded search recovers only 5\% of full adoption, against 42\% for one-hop search. Third, an inbox comprises four decisions (what to read, what to keep, what a sender pays, and what a reply may say), and the best rule for each depends on the traffic and the agent’s role. Multi-agent coordination under load requires an inbox layer designed for the traffic each agent receives.
\end{abstract}


\maketitle

\section{Introduction}
\label{sec:intro}


The frontier of large language model-based AI agent capabilities is achieved today, not with one agent, but with groups (or societies) of agents communicating and coordinating with each other~\cite{openai2026hf,openai2026navierstokes}. This raises the question of effective, efficient and scalable communication: how should societies of agents share information, knowledge and context with each other. 

Existing work focuses on delivery (sending the message or \emph{push}) as the core problem to solve. For example,~\cite{a2a2026,mcp2025,nlip2025,slim2026} address the problem at a protocol level for efficient message passing between groups of agents. Other classes of work~\cite{chopra2025ripple} study the information that should be carried in these messages to make coordination effective. This paper studies an orthogonal, less-explored and arguably equally important aspect of the scalable coordination problem in large LLM-agent societies: how should an agent decide which message to receive and reason on, and what to ignore (the \emph{pull} problem, or `what to read and respond to'). 

This, of course, is not a new problem framing. Human societies show evidence of constrained social contexts: humans can only effectively pay attention to a small, finite number of peers in the society~\cite{dunbar1993coevolution}. Similarly,
for a language-model agent, delivery (or sending a message) is cheap. The scarce resource is
what the agent reads \cite{simon1971designing}. Each read message costs tokens, and occupies finite context, so every new message displaces something already there. 


We study the \emph{pull} problem by introducing a new primitive: the agent inbox. Think of the agent inbox as being similar to a human email inbox in purpose (but not in implementation or design), in which messages wait outside
the context of an agent, before the agent's own rules decide what it reads and reacts to versus what is ignored. 
This makes the inbox a private decision with public consequences. Each agent sets its rules independently for handling the inbox, without senders being able to see them, yet those rules decide the effectiveness of society-level coordination. When requests are forwarded, they also decide whether anyone behind that agent is reached (§\ref{sec:method:multihop}). We therefore call the inbox a coordination primitive: a mechanism that every form of coordination in the society runs through. 

To this end, we study inboxes in simulated societies of up to 16,000 agents across a range of goals, from negotiating shared schedules, to forming groups of agents with a common need. In these societies, we study three questions: \emph{first}, how do push and pull hold up as an agent takes part in more coordination at once and receives more traffic, \emph{second}, is the inbox as a primitive a need-to-have or a good-to-have, and \emph{third}, what are some of the core design choices that need to be considered as we design a scalable coordination primitive for agent societies. 


Our contributions are the following:
\begin{enumerate}
  \item We frame agent communication as push or pull, and we show that push
  fails as an agent takes part in more coordination at once. Under push, an
  agent leading 20 negotiations with a full memory proposes options that were
  already rejected 97\% of the time, because it has dropped the oldest
  replies (\autoref{fig:organizer}).
  \item We show that an inbox is necessary in every shape of coordination we
    study. At a spam level where first come, first read misses every request,
  an inbox that first reads messages carrying a refundable deposit still finds
  about 60\% of the agents a request needs, at every society size from 1,000
  to 16,000 (\autoref{sec:eval:topology}).
  \item We break an inbox into four decisions, which are what to read, what
  to keep, what a sender pays for attention, and what a reply may say. For
  each decision, we compare several rules and show that the best rule depends
  on the traffic and on the agent's role, so no single inbox design suits
  every society (\autoref{sec:eval:policies}, \autoref{sec:eval:dilution}).
\end{enumerate}

In short, how an agent decides what to read now shapes what the whole society can accomplish.

\subsection{Scope}
We do not propose a new inbox implementation or claim that any of the rules we test is optimal; we use deliberately simple rules to isolate their effects on coordination, and we study four consequential choices rather than every inbox design decision: what to read, what to keep, what to charge for attention, and what a reply may disclose. Our claim is that coordination under load requires agents to manage incoming attention, and that these choices have consequences beyond the individual agent. The specific mechanisms will evolve; our contribution is to establish the inbox as a coordination primitive and show why its design matters.

\section{Related Work}
\label{sec:related}

\paragraph{Coordination and communication cost.}
Multi-agent coordination research has largely studied the cost
of sending and routing messages. Contract Net and matchmakers
coordinate tasks and requests \cite{smith1980contract,kuokka1995matchmaking},
while coalition formation studies how agents find and negotiate
with partners \cite{tsvetovat2000customer,bagnoli1989provision,
sarne2003search}. KQML and FIPA define how agents communicate
and refuse requests \cite{finin1994kqml,fipa2002cal}.
Other work studies communication costs
\cite{endriss2005communication,procaccia2006communication},
scalable search and routing \cite{lv2002search,yang2002improving,
stoica2001chord,castro2002scribe,kleinberg2000navigation},
and incentives for forwarding \cite{kleinberg2005query,
bhattacharyya2024referral}. Recent methods limit messages
\cite{han2023sparsecomm}, allocate speaking turns
\cite{fan2026costeffective}, or price communication
\cite{harris2026inference}. These approaches primarily
manage what senders transmit. We study the recipient's
limited capacity to read and remember incoming messages.

\paragraph{Inbox rules and attention.}
The rules we study build on established ideas in queueing,
caching, and email management. Reading policies include
FIFO, gated service, and priority queues
\cite{harcholbalter2013performance}, while memory policies
relate to cache replacement
\cite{belady1966study,sleator1985amortized}.
Email research studies overload, filtering, and prioritization
\cite{whittaker1996email,dabbish2006email,
sahami1998bayesian,aberdeen2010learning}.
Pricing attention has also been studied through postage,
refundable deposits, and queueing tolls
\cite{dwork1992pricing,loder2006economic,naor1969regulation}.
Attention and metareasoning have known costs
\cite{horvitz2003models,russell1991principles}.
We do not propose new queueing or pricing rules. We study
how independently chosen inbox rules affect coordination
when agents have limited attention and memory.

\paragraph{LLM-agent systems.}
LLM-agent simulations study populations of thousands to
millions of agents \cite{park2023generative,park2024thousand,
yang2024oasis,chopra2025lpm}, while benchmarks examine
negotiation and communication failures
\cite{bansal2025magentic,zhu2025automated,cemri2025mast}.
Communication pruning reduces messages
\cite{zhang2025agentprune}, and selective communication
lets cooperative agents choose which peers to hear
\cite{bhatt2026uncap}. Protocols
\cite{anthropic2024mcp,google2025a2a} and registries
\cite{raskar2025nanda,muscariello2025agntcy} support
communication and discovery. Memory systems study what
individual agents retain or share
\cite{packer2023memgpt,zhang2025memsurvey,
chen2025ioa,kim2026matm}. Recent work also studies
selective message passing between reasoning threads
\cite{liu2026messagepassing}. CalBench evaluates meeting
scheduling among five agents without a binding memory
limit \cite{zou2026calbench}. Our focus is on societies
where agents face competing requests, limited reading
budgets, and bounded memory. We study how each recipient's
inbox rules affect which agents can coordinate, including
when requests must pass through several recipients.

\section{The Inbox}
\label{sec:system}

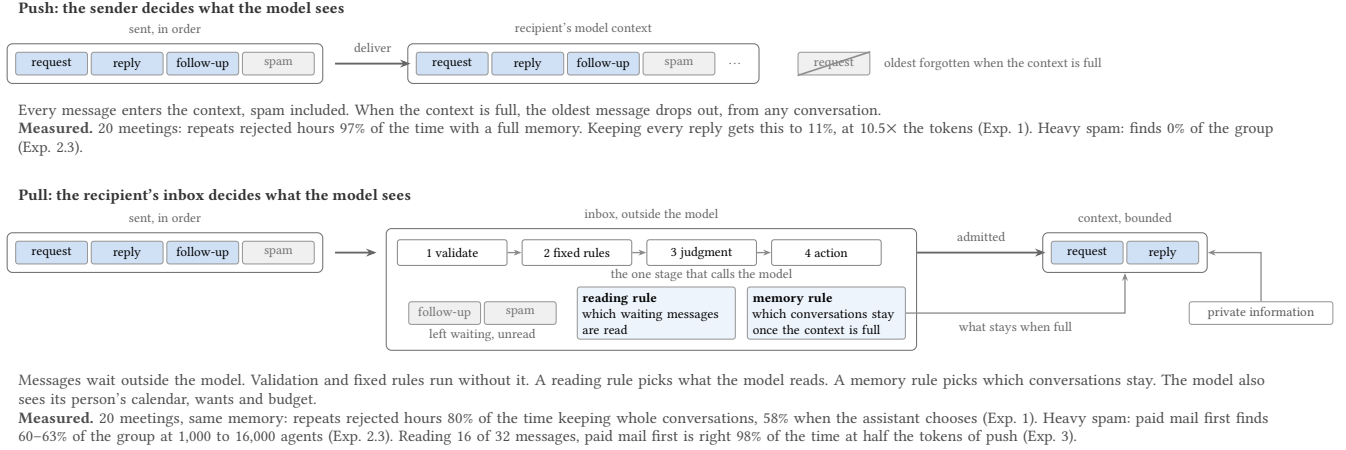
\begin{figure*}[t]
\centering
\begin{tikzpicture}[x=1cm, y=1cm, font=\scriptsize,
  tile/.style={draw=black!40, rounded corners=1pt, minimum width=0.95cm, minimum height=0.3cm, inner sep=1pt, font=\tiny},
  ta/.style={tile, fill=cbest!22}, ts/.style={tile, fill=black!6, text=black!55},
  stage/.style={draw=black!45, rounded corners=1pt, minimum width=1.45cm, minimum height=0.34cm, inner sep=1pt, font=\tiny, fill=white},
  box/.style={draw=black!55, rounded corners=2pt, inner sep=2pt},
  rule/.style={draw=black!40, rounded corners=1pt, fill=cbest!8, inner sep=2pt, font=\tiny, text width=1.95cm, align=left, minimum height=0.5cm},
  lab/.style={font=\scriptsize, text=black!80, anchor=west},
  capt/.style={font=\scriptsize, text=black!70, anchor=north west, text width=16.6cm},
  blab/.style={font=\tiny, text=black!60},
  arr/.style={-{Stealth[length=4pt]}, black!60, line width=0.7pt},
  farr/.style={-{Stealth[length=3pt]}, black!45, line width=0.5pt}]
\node[lab] at (0,4.25) {\textbf{Push: the sender decides what the model sees}};
\foreach \i/\s/\n in {0/ta/request,1/ta/reply,2/ta/follow-up,3/ts/spam}
  \node[\s] (p\n) at (0.55+1.0*\i,3.5) {\n};
\node[box, fit=(prequest)(pspam), inner sep=3pt, label={[blab]above:sent, in order}] {};
\draw[arr] (4.3,3.5) -- node[above, blab, yshift=1pt] {deliver} (5.3,3.5);
\foreach \i/\s/\n in {0/ta/request,1/ta/reply,2/ta/follow-up,3/ts/spam}
  \node[\s] (c\n) at (5.85+1.0*\i,3.5) {\n};
\node[blab] (cdots) at (9.6,3.5) {\ldots};
\node[box, fit=(crequest)(cspam)(cdots), inner sep=3pt, label={[blab]above:recipient's model context}] {};
\node[ts] (drop) at (10.9,3.5) {request};
\draw[black!45, line width=0.6pt] (drop.south west) -- (drop.north east);
\node[blab, anchor=west] at (11.45,3.5) {oldest forgotten when the context is full};
\node[capt] at (0,3.08) {Every message enters the context, spam included. When the context is full, the oldest message drops out, from any conversation.\\
\textbf{Measured.} 20 meetings: repeats rejected hours 97\% of the time with a full memory. Keeping every reply gets this to 11\%, at 10.5$\times$ the tokens (Exp.~1). Heavy spam: finds 0\% of the group (Exp.~2.3).};
\node[lab] at (0,1.75) {\textbf{Pull: the recipient's inbox decides what the model sees}};
\foreach \i/\s/\n in {0/ta/request,1/ta/reply,2/ta/follow-up,3/ts/spam}
  \node[\s] (r\n) at (0.55+1.0*\i,1.0) {\n};
\node[box, fit=(rrequest)(rspam), inner sep=3pt, label={[blab]above:sent, in order}] {};
\draw[arr] (4.3,1.0) -- (4.85,1.0);
\node[stage] (s1) at (5.85,1.0) {1 validate};
\node[stage] (s2) at (7.5,1.0) {2 fixed rules};
\node[stage] (s3) at (9.15,1.0) {3 judgment};
\node[stage] (s4) at (10.8,1.0) {4 action};
\draw[farr] (s1) -- (s2); \draw[farr] (s2) -- (s3); \draw[farr] (s3) -- (s4);
\node[blab, anchor=north, inner sep=1pt] at (s3.south) {the one stage that calls the model};
\node[ts] (wfollow) at (5.75,0.2) {follow-up};
\node[ts] (wspam) at (6.75,0.2) {spam};
\node[blab, anchor=north, inner sep=1pt] at (6.25,0.03) {left waiting, unread};
\node[rule] (rr) at (8.55,0.2) {\textbf{reading rule}\\which waiting messages are read};
\node[rule] (mr) at (10.8,0.2) {\textbf{memory rule}\\which conversations stay once the context is full};
\node[box, fit=(s1)(s4)(wfollow)(mr), inner sep=4pt, label={[blab]above:inbox, outside the model}] (inbox) {};
\draw[arr] (inbox.east |- s4) -- node[above, blab, yshift=1pt] {admitted} (13.7,1.0);
\foreach \i/\s/\n in {0/ta/request,1/ta/reply}
  \node[\s] (e\n) at (14.25+1.0*\i,1.0) {\n};
\node[box, fit=(erequest)(ereply), inner sep=3pt, label={[blab]above:context, bounded}] (ctx) {};
\draw[farr] (mr.east) -| node[pos=0.25, below, blab] {what stays when full} (ctx.south);
\node[tile, fill=white, text=black!65, minimum width=1.9cm] (priv) at (16.55,0.2) {private information};
\draw[farr] (priv.north) |- (ctx.east);
\node[capt] at (0,-0.5) {Messages wait outside the model. Validation and fixed rules run without it. A reading rule picks what the model reads. A memory rule picks which conversations stay. The model also sees its person's calendar, wants and budget.\\
\textbf{Measured.} 20 meetings, same memory: repeats rejected hours 80\% of the time keeping whole conversations, 58\% when the assistant chooses (Exp.~1). Heavy spam: paid mail first finds 60--63\% of the group at 1{,}000 to 16{,}000 agents (Exp.~2.3). Reading 16 of 32 messages, paid mail first is right 98\% of the time at half the tokens of push (Exp.~3).};
\end{tikzpicture}
\caption{Push and pull. Under push, every message enters the context in
arrival order and the oldest is forgotten when it fills. Under pull,
messages wait in an inbox: a reading rule picks what is read, a memory
rule what stays, and only judgment calls the model.}
\Description{Two-row diagram. Top row, push: a request, a reply, a follow-up and a spam message go straight into the recipient's context in arrival order, and the oldest are dropped when the context is full. Bottom row, pull: the same four messages wait in an inbox. A reading rule picks which are read, a memory rule picks which conversations stay in context, and only the judgment step calls the model. Headline numbers from Experiments 1 to 3 appear under each row.}
\label{fig:architecture}
\end{figure*}

An agent may receive more messages than it can read, and its
context cannot retain every earlier exchange. Two choices
follow: which messages to read now, and what information to
keep for later use. Similar problems arise in queueing and
caching \cite{harcholbalter2013performance,sleator1985amortized}.
Coordination introduces two further choices. When several
senders compete for attention, the recipient may allow them
to post a refundable stake for priority, following earlier
work on attention bonds \cite{loder2006economic}. When
responding, the agent must decide what information to share.
A yes or no answer may be insufficient for coordination,
while revealing the private information behind a decision
may disclose more than necessary \cite{zou2026calbench}.
We study these four choices as important parts of inbox
design.

\paragraph{Inbox model.}
An inbox holds incoming messages outside the model context
(Figure~\ref{fig:architecture}). Messages first pass
validation and fixed rules about senders and categories,
without calling the model. The remaining messages wait
until the agent selects which ones to read.

At step $t$, agent $i$ has a queue $Q_i^t$ of waiting
messages and a context $K_i^t$ containing retained
information. The agent can read at most $b$ messages
per step, and its context holds at most $C$ tokens.
We represent the inbox as $I_i=(R_i,M_i,d_i,V_i)$.

The \emph{reading rule} $R_i$ selects which waiting
messages to read,
\[
A_i^t=R_i(Q_i^t), \qquad |A_i^t|\leq b.
\]
The \emph{memory rule} $M_i$ selects what to retain
after reading,
\[
K_i^{t+1}=M_i(K_i^t\cup A_i^t),
\qquad \ell(K_i^{t+1})\leq C,
\]
where $\ell(\cdot)$ counts tokens. Reading and memory
are separate decisions. A message may be read now but
forgotten before a later decision requires it.

The \emph{sender stake} $d_i\geq 0$ is a refundable
deposit set by the recipient for reading priority.
A sender can attach the deposit to move its message
ahead of others. The stake affects reading order,
but does not guarantee acceptance or a useful reply.
The \emph{reply vocabulary} $V_i$ limits what the agent
may disclose. Replies range from a simple yes or no
to conditions, exact limits, or private information.
A reply with more detail can help other agents decide
what to request next, but may also reveal information
the recipient would prefer to keep private.

The model judges selected messages using retained
context and private information. It then decides
whether and how to respond. Only selected messages
are passed to the model for judgment. When a reading
or memory rule requires another model call, we count
its additional token cost.

\paragraph{Rules.}
Table~\ref{tab:inbox} lists the rules we compare.
The reading rules differ in how they select messages,
and the memory rules differ in what they preserve.
Paid mail first is the only reading rule that uses
a stake. The deposit is refunded when a message is
read and answered, and forfeited otherwise. We also
test what happens when unwanted messages carry deposits.
The rules are representative rather than exhaustive,
and we do not claim that any rule is optimal.

Under \emph{push}, messages are read in arrival order,
the oldest messages are forgotten when context fills,
and no stake is required. Under \emph{pull} (the regime this work studies and proposesfig:architecture), messages
wait outside the context, and the recipient chooses
what to read and retain. In our simulations, every
agent can be contacted only through its inbox.
A request must be read before it expires for the
agent to consider it. Each agent sets its rules
independently, but unread requests cannot contribute
to coordination. When requests are forwarded, each
inbox also determines whether other agents further
along the path can be reached.

\begin{table}[t]
\caption{The four inbox decisions and the rules compared.
Push uses first come, first read; forget the oldest;
no stake; and unrestricted replies.}
\label{tab:inbox}
\small
\begin{tabular}{@{}p{0.39\columnwidth}p{0.56\columnwidth}@{}}
\toprule
\multicolumn{2}{@{}l}{\textbf{Reading} $R_i$: what to read} \\
First come, first read & Read in arrival order \\
Check in batches & Wait several steps, then clear the queue \\
Sort into folders & Prioritize by category; ignore junk \\
Newest first & Read newest messages first \\
Paid mail first & Read messages with stakes first \\
Closest match & Rank by similarity to published interests \\
Agent chooses & Model selects from one line per message \\
\midrule
\multicolumn{2}{@{}l}{\textbf{Memory} $M_i$: what to retain} \\
Forget the oldest & Drop oldest messages \\
Summarize the old & Keep latest reply per sender from older half \\
Keep whole conversations & Drop least recently active conversations \\
Agent chooses & Model selects conversations to retain \\
Promise list & Keep promised hours for each open meeting \\
\midrule
\multicolumn{2}{@{}l}{\textbf{Sender stake} $d_i$: priority for a deposit} \\
No stake & No paid priority \\
Refundable stake & Return when read and answered; otherwise forfeit \\
\midrule
\multicolumn{2}{@{}l}{\textbf{Reply} $V_i$: what may be disclosed} \\
Bare & Yes or no \\
Coarse & Yes with one condition \\
Conditional & Yes with a limit from private information \\
Full & Disclose private information \\
\bottomrule
\end{tabular}
\end{table}

\section{Experiments}
\label{sec:method}
\label{sec:method:experiments}
\label{sec:results}

We study three questions. First, why does push become costly
or unreliable as agents participate in more interactions?
Second, do these limitations affect coordination across
different agent roles, team sizes, society sizes, and networks?
Third, which inbox rules work under different reading limits
and traffic? Additional experiments test the effects of
discovery, adoption, competing requests, and adversarial
senders (Table~\ref{tab:design}).

\paragraph{Coordination tasks.}
We use two tasks with measurable outcomes. In
\emph{meeting scheduling}, agents negotiate shared times
over several rounds, using private calendars and earlier
commitments. Each agent may participate in several
negotiations at once. In \emph{group formation}, a requester
must find enough agents with a matching private interest
before its request expires. We test both direct and forwarded
communication. While specific, these tasks let us study broader coordination
problems involving limited attention, bounded memory,
and dependence on other agents.

\paragraph{Agent decisions and comparisons.}
The simulation supplies messages, private information,
network structure, and correct outcomes. An LLM makes
decisions using only the information available to its
agent, without access to other private information or
the correct answer. Identical inputs may share a model
output to reduce simulation cost. We use gpt-5.6-terra
unless stated otherwise, with additional tests using
gpt-5.4-mini and gpt-5.6-sol.

Where possible, rules face the same decisions under
equal reading and memory limits. Controls retain all
information or retrieve exactly what a decision needs.
We test both manageable and heavy traffic, since
failure under extreme overload alone is insufficient
to establish the value of an inbox.

\paragraph{Measures.}
For scheduling, we measure repeated proposals and
conflicting commitments. For group formation, we
measure the share of relevant agents found and the
groups formed. We also record whether the relevant
message is read, the correctness of the response,
and token cost.

\begin{table*}[t]
\caption{Summary of experiments. Reading and memory rules
are defined in Table~\ref{tab:inbox}. Experiments 1, 2.1,
2.2, and 3 use 100 decisions per setting.}
\label{tab:design}
\small
\begin{tabular}{@{}p{0.18\textwidth}p{0.61\textwidth}p{0.15\textwidth}@{}}
\toprule
Experiment & Conditions compared & Measures \\
\midrule
1. Concurrent negotiations
& 2, 5, 10, 20 meetings; memory rules at 375, 750,
1{,}500 tokens; full-memory and best-case controls;
five invitees, two rounds, five societies
& Repeated proposals; tokens \\

2.1 Team size
& 2, 5, 8, 12 invitees; memory for one or two
conversations; concurrent meetings; five societies
& Repeated proposals \\

2.2 Agent role
& 2, 5, 10, 20 invitations; memory rules including
a promise list; five societies
& Conflicting commitments \\

2.3 Society size
& 1{,}000, 4{,}000, 16{,}000 agents; push and paid
mail first; three search methods; 24 matching agents,
group of ten; heavy traffic, ten societies
& Share found; groups formed \\

2.4 Network structure
& 50, 250, 1{,}005 agents; direct and forwarded search;
inbox rules, traffic, and adoption; 20 requests,
three societies per size
& Share found; forwarding losses \\

3. Reading rules
& 4, 8, 16 reads from 32 messages; reading rules
and reply conditions
& Target read; correct replies; tokens \\

Supporting experiments
& Partial adoption, perfect matching, competing traffic,
sender stakes, and additional context
& Reach; accuracy; tokens \\
\bottomrule
\end{tabular}
\end{table*}

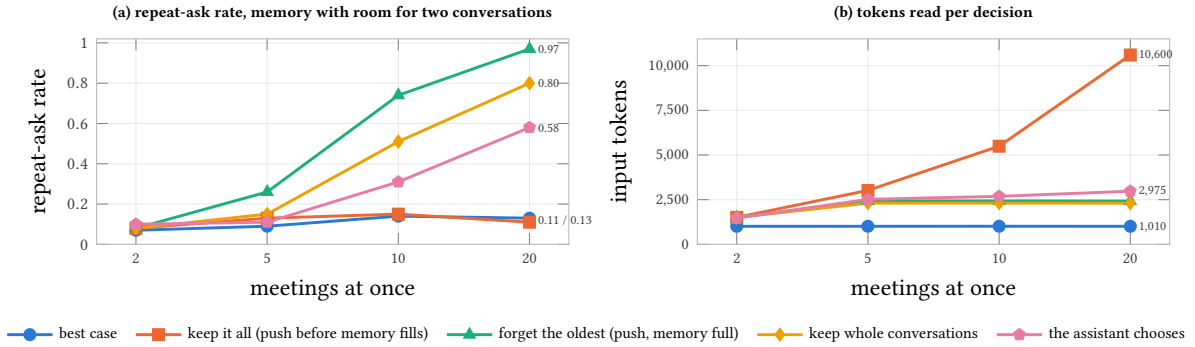
\begin{figure*}[t]
\centering
\begin{tikzpicture}
\begin{groupplot}[group style={group size=2 by 1, horizontal sep=1.7cm},
  paperaxis, width=0.44\textwidth, height=4.3cm,
  symbolic x coords={2,5,10,20}, xtick=data, xlabel={meetings at once}]
\nextgroupplot[title={(a) repeat-ask rate, memory with room for two conversations},
  ylabel={repeat-ask rate},
  ymin=0, ymax=1.02, legend to name=legx1, legend columns=-1]
\addplot[sbest] coordinates {(2,0.07)(5,0.09)(10,0.14)(20,0.13)}; \addlegendentry{best case}
\addplot[skeep] coordinates {(2,0.08)(5,0.13)(10,0.15)(20,0.11)}; \addlegendentry{keep it all (push before memory fills)}
\addplot[sforget] coordinates {(2,0.08)(5,0.26)(10,0.74)(20,0.97)}; \addlegendentry{forget the oldest (push, memory full)}
\addplot[swhole] coordinates {(2,0.08)(5,0.15)(10,0.51)(20,0.80)}; \addlegendentry{keep whole conversations}
\addplot[sassist] coordinates {(2,0.10)(5,0.11)(10,0.31)(20,0.58)}; \addlegendentry{the assistant chooses}
\node[endlab] at (axis cs:20,0.97) {0.97};
\node[endlab] at (axis cs:20,0.80) {0.80};
\node[endlab] at (axis cs:20,0.58) {0.58};
\node[endlab] at (axis cs:20,0.12) {0.11 / 0.13};
\nextgroupplot[title={(b) tokens read per decision}, ylabel={input tokens},
  ymin=0, ymax=11500, scaled y ticks=false, ytick={0,2500,5000,7500,10000}, yticklabels={0,{2,500},{5,000},{7,500},{10,000}}]
\addplot[sbest] coordinates {(2,1010)(5,1011)(10,1013)(20,1010)};
\addplot[skeep] coordinates {(2,1509)(5,3020)(10,5493)(20,10596)};
\addplot[sforget] coordinates {(2,1509)(5,2421)(10,2432)(20,2422)};
\addplot[swhole] coordinates {(2,1509)(5,2307)(10,2309)(20,2306)};
\addplot[sassist] coordinates {(2,1469)(5,2520)(10,2687)(20,2975)};
\node[endlab] at (axis cs:20,10596) {10,600};
\node[endlab] at (axis cs:20,2975) {2,975};
\node[endlab] at (axis cs:20,1010) {1,010};
\end{groupplot}
\end{tikzpicture}\\[2pt]
\pgfplotslegendfromname{legx1}
\caption{An organizer running 2 to 20 meetings at once;
five invitees, 100 decisions per cell, Wilson 95\%
intervals~\cite{wilson1927probable} $\le 0.10$. (a) Keeping every reply
matches the best case; with a full memory, forgetting the oldest (push)
collapses and the inbox rules collapse less. (b) Keeping every reply
costs 1.5--10.5$\times$ the best case's tokens.}
\Description{Two line charts over 2, 5, 10 and 20 meetings at once, one line per memory rule. Left: share of proposed hours already rejected. Keeping it all tracks the best case at every load. Forgetting the oldest rises to 0.97 at twenty meetings. Keeping whole conversations and letting the assistant choose stay lower. Right: input tokens per decision. Keeping it all grows with load to about ten times the best case, while the two inbox rules read a fifth to a quarter as much.}
\label{fig:organizer}
\end{figure*}

\subsection{Concurrent Coordination with Limited Memory}
\label{sec:method:organizer}
\label{sec:why}

Under push, incoming messages enter the model context in arrival
order. As an agent participates in more concurrent interactions,
it must either read a growing history or discard older messages.
We test whether errors arise because the model cannot use a
long history, because reading it becomes expensive, or because
important information is forgotten.

We use meeting scheduling because each negotiation produces
a history of proposals and replies that later decisions must
respect. An agent organizes 2 to 20 meetings concurrently.
After two rounds of replies, we ask it to propose new times
for one randomly selected meeting that remains open and has
a feasible solution. We compare several ways of managing the
same history. \emph{Keep it all} retains every reply, while
\emph{forget the oldest} drops messages in arrival order.
Other rules summarize older messages, keep whole conversations,
or let the model choose which conversations to retain.
The \emph{best case} retrieves exactly the conversation needed
for the next decision, providing a reference with perfect recall.

All rules face the same decisions across five simulated
societies. The bounded-memory rules have the same context
limits, chosen relative to the size of a conversation
(Table~\ref{tab:design}). These limits favor keeping whole
conversations. Each rule must also decide what to retain
before knowing which meeting will be selected. Only the
best-case control receives the relevant conversation directly.
We measure how often the agent proposes times that have
already been rejected, and the tokens required per decision.

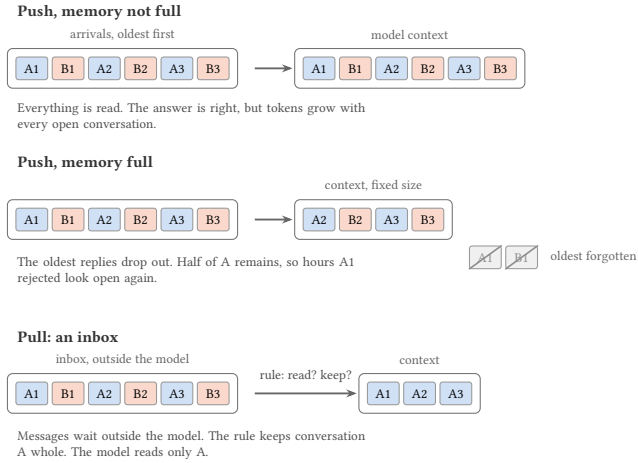
\begin{figure}[t]
\centering
\begin{tikzpicture}[x=1cm, y=1cm, font=\scriptsize,
  tile/.style={draw=black!40, rounded corners=1pt, minimum width=0.42cm, minimum height=0.3cm, inner sep=1pt, font=\tiny},
  ta/.style={tile, fill=cbest!22}, tb/.style={tile, fill=ckeep!25},
  box/.style={draw=black!55, rounded corners=2pt, inner sep=2pt},
  lab/.style={font=\scriptsize, text=black!80, anchor=west},
  capt/.style={font=\tiny, text=black!70, anchor=north west, text width=4.6cm},
  arr/.style={-{Stealth[length=4pt]}, black!60, line width=0.7pt}]
\node[lab] at (0,3.75) {\textbf{Push, memory not full}};
\foreach \i/\s/\n in {0/ta/A1,1/tb/B1,2/ta/A2,3/tb/B2,4/ta/A3,5/tb/B3}
  \node[\s] (p\n) at (0.3+0.48*\i,3.0) {\n};
\node[box, fit=(pA1)(pB3), inner sep=3pt, label={[font=\tiny, text=black!60]above:arrivals, oldest first}] {};
\draw[arr] (3.25,3.0) -- (3.75,3.0);
\foreach \i/\s/\n in {0/ta/A1,1/tb/B1,2/ta/A2,3/tb/B2,4/ta/A3,5/tb/B3}
  \node[\s] (c\n) at (4.1+0.48*\i,3.0) {\n};
\node[box, fit=(cA1)(cB3), inner sep=3pt, label={[font=\tiny, text=black!60]above:model context}] {};
\node[capt] at (0,2.65) {Everything is read. The answer is right, but tokens grow with every open conversation.};
\node[lab] at (0,1.75) {\textbf{Push, memory full}};
\foreach \i/\s/\n in {0/ta/A1,1/tb/B1,2/ta/A2,3/tb/B2,4/ta/A3,5/tb/B3}
  \node[\s] (q\n) at (0.3+0.48*\i,1.0) {\n};
\node[box, fit=(qA1)(qB3), inner sep=3pt] {};
\draw[arr] (3.25,1.0) -- (3.75,1.0);
\foreach \i/\s/\n in {0/ta/A2,1/tb/B2,2/ta/A3,3/tb/B3}
  \node[\s] (d\n) at (4.1+0.48*\i,1.0) {\n};
\node[box, fit=(dA2)(dB3), inner sep=3pt, label={[font=\tiny, text=black!60]above:context, fixed size}] {};
\node[tile, fill=black!6, text=black!45] (dropA1) at (6.3,0.5) {A1};
\node[tile, fill=black!6, text=black!45] (dropB1) at (6.78,0.5) {B1};
\draw[black!45, line width=0.6pt] (dropA1.south west) -- (dropA1.north east) (dropB1.south west) -- (dropB1.north east);
\node[font=\tiny, text=black!60, anchor=west] at (7.05,0.5) {oldest forgotten};
\node[capt] at (0,0.62) {The oldest replies drop out. Half of A remains, so hours A1 rejected look open again.};
\node[lab] at (0,-0.55) {\textbf{Pull: an inbox}};
\foreach \i/\s/\n in {0/ta/A1,1/tb/B1,2/ta/A2,3/tb/B2,4/ta/A3,5/tb/B3}
  \node[\s] (r\n) at (0.3+0.48*\i,-1.3) {\n};
\node[box, fit=(rA1)(rB3), inner sep=3pt, label={[font=\tiny, text=black!60]above:inbox, outside the model}] {};
\draw[arr] (3.25,-1.3) -- node[above, font=\tiny, text=black!70, yshift=1pt] {rule: read? keep?} (4.6,-1.3);
\foreach \i/\s/\n in {0/ta/A1,1/ta/A2,2/ta/A3}
  \node[\s] (e\n) at (4.95+0.48*\i,-1.3) {\n};
\node[box, fit=(eA1)(eA3), inner sep=3pt, label={[font=\tiny, text=black!60]above:context}] {};
\node[capt] at (0,-1.68) {Messages wait outside the model. The rule keeps conversation A whole. The model reads only A.};
\end{tikzpicture}
\caption{Why forgetting the oldest fails. Six replies from two
negotiations, A and B. Push with a full context drops A1 and B1 by age,
so the organizer re-proposes hours A1 had rejected. A pull inbox keeps
A whole outside the model and reads only A.}
\Description{Diagram of three ways to hold six replies from two conversations, A and B. Push with memory not yet full keeps all six. Push with a full memory keeps only the newest replies and splits both conversations. A pull inbox stores whole conversations outside the model and reads only the conversation the decision needs.}
\label{fig:pushpull}
\end{figure}

Figure~\ref{fig:organizer}a shows that retaining the full
history performs almost as well as perfect retrieval at
every tested load. The model can therefore use a long history
without substantial loss of accuracy in this setting.
However, reading that history becomes increasingly expensive.
At 20 concurrent meetings, keeping everything requires
10.5 times the tokens of perfect retrieval
(Figure~\ref{fig:organizer}b).

With limited memory, the problem is different. At 20
meetings, forgetting the oldest messages causes 97\% of proposals to repeat previously rejected times.  Figure~\ref{fig:pushpull} illustrates why. Removing messages by age can leave part of a conversation in memory while discarding earlier rejections. When the required conversation remains intact, the agent performs as well as with perfect retrieval. When earlier replies are missing, the agent may propose times that were already rejected. Keeping whole conversations avoids incomplete histories, but may discard the conversation needed next. Letting the model choose what to retain performs better. At the same memory limit, the repeat-ask rates are 80\% and 58\%, respectively, compared with about 11\% when the entire history is available. Both inbox rules
also read substantially fewer tokens than keeping
everything. The share of feasible proposals follows the same pattern.

The same ordering holds across the five societies.
The results also depend on the model instructions, and an earlier difference between full history and perfect retrieval disappeared after correcting the instructions. The main finding is therefore about the cost of retaining history and the consequences of losing earlier replies, rather than a fixed ranking of memory rules.

At two concurrent meetings, all memory rules perform
similarly. CalBench \cite{zou2026calbench} studies scheduling
among five agents without a binding memory limit.
Our experiment examines what changes as agents participate
in more concurrent negotiations and earlier replies can
no longer all be retained.

\takeaway{Takeaway 1: Scaling coordination requires pull}{
As agents participate in more concurrent interactions,
push requires either reading an increasingly large history
or forgetting information needed for later decisions.
Pull separates receiving messages from deciding what to
read and retain, allowing agents to manage attention and
memory as coordination grows.
}

\subsection{Why Coordination Needs an Inbox}
\label{sec:method:shapes}
\label{sec:eval:topology}

Experiment~1 showed that push becomes costly or unreliable
as an agent participates in more concurrent interactions.
We now test whether the need for an inbox extends to other
forms of coordination. We vary team size, agent role, society
size, and network structure to examine where the limitations
of push arise and how they affect other agents. The goal is
to understand whether managing incoming messages is necessary
for coordination across a society, rather than only for
managing the memory of one agent.

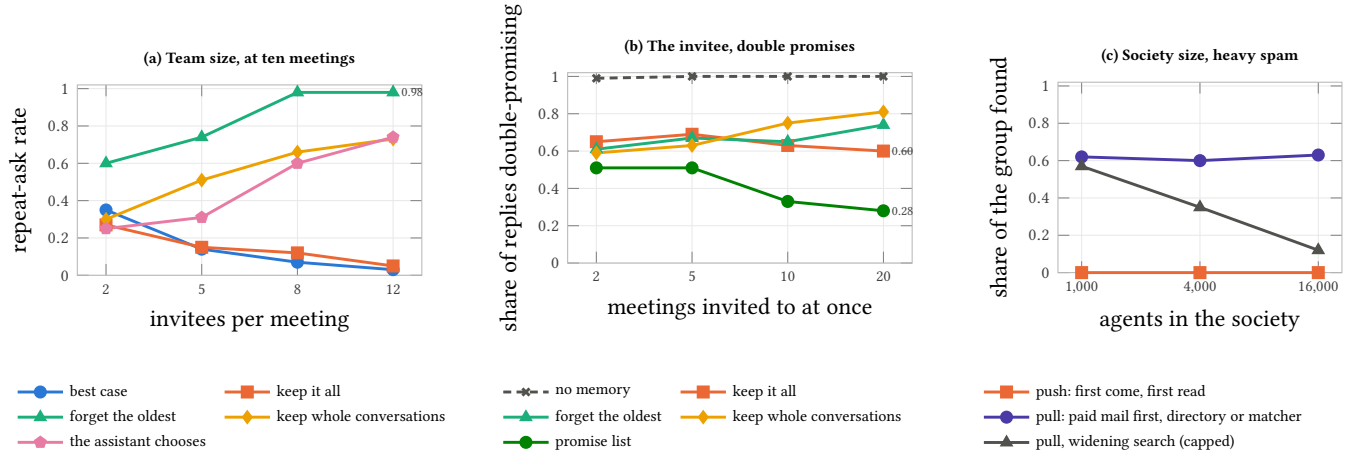
\begin{figure*}[t]
\centering
\begin{tikzpicture}
\begin{axis}[paperaxis, width=0.345\textwidth, title={(a) Team size, at ten meetings},
  symbolic x coords={2,5,8,12}, xtick=data, xlabel={invitees per meeting}, ylabel={repeat-ask rate}, ymin=0, ymax=1.02,
  legend to name=legx2a, legend columns=2, legend cell align=left]
\addplot[sbest] coordinates {(2,0.35)(5,0.14)(8,0.07)(12,0.03)}; \addlegendentry{best case}
\addplot[skeep] coordinates {(2,0.27)(5,0.15)(8,0.12)(12,0.05)}; \addlegendentry{keep it all}
\addplot[sforget] coordinates {(2,0.60)(5,0.74)(8,0.98)(12,0.98)}; \addlegendentry{forget the oldest}
\addplot[swhole] coordinates {(2,0.30)(5,0.51)(8,0.66)(12,0.73)}; \addlegendentry{keep whole conversations}
\addplot[sassist] coordinates {(2,0.25)(5,0.31)(8,0.60)(12,0.74)}; \addlegendentry{the assistant chooses}
\node[endlab] at (axis cs:12,0.98) {0.98};
\end{axis}
\end{tikzpicture}\hfill
\begin{tikzpicture}
\begin{axis}[paperaxis, width=0.345\textwidth, title={(b) The invitee, double promises},
  symbolic x coords={2,5,10,20}, xtick=data, xlabel={meetings invited to at once}, ylabel={share of replies double-promising}, ymin=0, ymax=1.02,
  legend to name=legx2b, legend columns=2, legend cell align=left]
\addplot[snone] coordinates {(2,0.99)(5,1.00)(10,1.00)(20,1.00)}; \addlegendentry{no memory}
\addplot[skeep] coordinates {(2,0.65)(5,0.69)(10,0.63)(20,0.60)}; \addlegendentry{keep it all}
\addplot[sforget] coordinates {(2,0.61)(5,0.67)(10,0.65)(20,0.74)}; \addlegendentry{forget the oldest}
\addplot[swhole] coordinates {(2,0.59)(5,0.63)(10,0.75)(20,0.81)}; \addlegendentry{keep whole conversations}
\addplot[spromise] coordinates {(2,0.51)(5,0.51)(10,0.33)(20,0.28)}; \addlegendentry{promise list}
\node[endlab] at (axis cs:20,0.60) {0.60};
\node[endlab] at (axis cs:20,0.28) {0.28};
\end{axis}
\end{tikzpicture}\hfill
\begin{tikzpicture}
\begin{axis}[paperaxis, width=0.3\textwidth, title={(c) Society size, heavy spam},
  symbolic x coords={1000,4000,16000}, xtick=data, xticklabels={{1,000},{4,000},{16,000}}, xlabel={agents in the society}, ylabel={share of the group found}, ymin=0, ymax=1.02,
  legend to name=legx2c, legend columns=1, legend cell align=left]
\addplot[skeep] coordinates {(1000,0.00)(4000,0.00)(16000,0.00)}; \addlegendentry{push: first come, first read}
\addplot[sdir] coordinates {(1000,0.62)(4000,0.60)(16000,0.63)}; \addlegendentry{pull: paid mail first, directory or matcher}
\addplot[swide] coordinates {(1000,0.57)(4000,0.35)(16000,0.12)}; \addlegendentry{pull, widening search (capped)}
\end{axis}
\end{tikzpicture}\\[2pt]
\makebox[0.345\textwidth][c]{\raisebox{-\height}{\pgfplotslegendfromname{legx2a}}}\hfill
\makebox[0.345\textwidth][c]{\raisebox{-\height}{\pgfplotslegendfromname{legx2b}}}\hfill
\makebox[0.3\textwidth][c]{\raisebox{-\height}{\pgfplotslegendfromname{legx2c}}}
\caption{Push degrades with each agent's load in every
shape of coordination. (a) Team size, ten meetings: keeping every reply
stays within 0.05 of the best case, so open negotiations, not team
size, break push. (b) The invitee double-promises even with every
promise in view; only the promise list improves with load. (c) Society
size, heavy spam: push finds no one, pull about 0.6 of the group at
every size.}
\Description{Three line charts. (a) Repeat-ask rate against invitees per meeting, from 2 to 12, at ten meetings: keeping it all stays within 0.05 of the best case. (b) Share of replies that promise an hour twice against meetings invited to at once, from 2 to 20: nearly always with no memory, 0.60 to 0.69 with everything in view, and only the promise list falls with load. (c) Share of the group found against society size, from 1,000 to 16,000 agents, under heavy spam: push finds nobody, pull through the directory or matcher finds about 0.6 at every size, and the capped widening search falls to 0.12.}
\label{fig:shapes}
\end{figure*}

\subsubsection{Team size}
We vary team size from two to twelve participants to test
whether the number of participants, rather than concurrent
interactions, causes push to fail. Memory is scaled to hold
one or two complete conversations. We evaluate meetings
that remain open and feasible after two rounds, and record
how often such meetings occur.

Retaining all replies remains within 0.05 of perfect
retrieval across team sizes, but costs three to eight
times as many tokens (Figure~\ref{fig:shapes}a).
The bounded-memory rules generally follow the ordering
in Experiment~1, although larger teams can change which
rule works best. With five concurrent meetings and twelve
participants, keeping whole conversations performs worse
than forgetting the oldest messages. Larger teams also
leave fewer feasible meetings to evaluate, which affects
the measured error rates.

\conclusion{Conclusion 2.1: Group size and concurrent interactions}{
The main limitation of push comes from managing concurrent
interactions, rather than group size alone. Larger groups
increase reading costs and can change which memory rule
works best.
}

\subsubsection{Agent role and memory}

Different roles require different information to be retained.
We consider an agent receiving 2 to 20 concurrent invitations.
Each reply must respect earlier commitments. We compare
full memory, no memory, bounded-memory rules, and a compact
\emph{promise list}. Every evaluated case includes a
previously committed time and an available alternative.
Earlier commitments are consistent, so any conflicting
reply introduces a new error.

Even with all earlier replies available, the model offers
committed times in 60\% to 69\% of cases
(Figure~\ref{fig:shapes}b). The model therefore makes errors
even when no information is missing. Forgetting messages
makes the problem worse. Keeping whole conversations,
which helped the organizer, performs poorly because
commitments are spread across conversations. A promise
list reduces conflicting replies to 28\% at 20 invitations,
although errors remain.

\conclusion{Conclusion 2.2: Memory depends on the agent role}{
Different roles require different information to be retained.
Keeping complete conversations helps an organizer, while a
participant needs commitments from across conversations.
Even complete history does not guarantee correct decisions.
}

\subsubsection{Society size}

We test whether inbox performance changes with society
size when the traffic received by each agent remains
comparable. In societies of 1{,}000 to 16{,}000 agents,
24 agents share a private interest, and a requester
must find ten of them. We compare push with paid mail
first, using widening search, a topic directory, or
a matcher that knows the relevant agents.

We deliberately choose enough unwanted traffic to
overwhelm arrival-order reading. Push therefore reads
none of the relevant requests by construction of the
stress condition. Paid mail first, using the directory
or matcher, finds about 60\% of relevant agents at
every society size (Figure~\ref{fig:shapes}c), forming
groups in roughly half the societies. Widening search
performs worse as society size grows because it
contacts at most 1{,}000 agents.

\conclusion{Conclusion 2.3: Discovery does not guarantee attention}{
Finding the correct agents does not ensure that they read
a request. Under heavy traffic, the reading rules of
individual recipients determine which requests are
considered, even when discovery is accurate.
}

\subsubsection{Forwarding across a contact network}
\label{sec:method:multihop}
\label{sec:eval:referral}

We next test whether inbox failures affect coordination
when requests pass through several agents. We use the
\texttt{email-Eu-core} contact graph from SNAP
\cite{leskovec2014snap,leskovec2007graph,yin2017local},
with 1{,}005 members and 24{,}929 directed connections,
and samples of 50 and 250 members
\cite{goodman1961snowball}. Matching interests are
grouped by department.

In network search, a requester contacts four agents,
and each recipient may forward to two contacts, for
up to four or five hops. Directory search contacts
the best matches directly, with a comparable limit.
We compare reading rules with and without unwanted
traffic, across 20 requests and three societies per
size. We also vary the fraction of agents using
paid mail first.

\begin{figure}[t]
  \centering
  \includegraphics[width=\columnwidth]{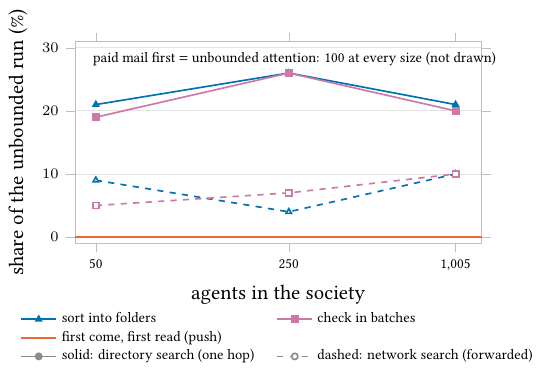}
  \caption{Inbox failures accumulate across agents.
  Group formation on \texttt{email-Eu-core} under heavy unwanted
  traffic, relative to unbounded attention. Solid: direct directory
  search. Dashed: forwarded through contacts. Paid mail first matches
  the control (not drawn); push finds no one; folders and batches lose
  far more when forwarded. Mean of three societies.}
  \Description{Share of the target group found relative
  to unbounded attention, across networks of 50, 250,
  and 1005 agents. Direct search performs better than
  forwarded search for folder sorting and batch reading.
  Arrival-order reading finds no agents, while paid
  mail first matches unbounded attention.}
  \label{fig:multihop}
\end{figure}

Inbox failures have larger effects when requests
must be forwarded (Figure~\ref{fig:multihop}).
Under heavy unwanted traffic, push finds no agents,
while paid mail first matches unbounded attention
because only genuine requests carry stakes. Sorting
into folders preserves 21\% to 26\% of the unbounded
result under direct search, but only 4\% to 10\%
when requests are forwarded. The losses persist
without unwanted traffic because the model frequently
classifies genuine requests as junk. A request missed
at one inbox cannot reach agents farther along
the same path.

\begin{table}[t]
\caption{Share of the target group found when a fraction
$f$ of agents use paid mail first and the rest use
arrival-order reading. Values are relative to full
adoption ($f=1$), under heavy unwanted traffic.
Mean of three societies. The last column gives the
mean number of hops under full adoption.}
\label{tab:adoption}
\small
\begin{tabular}{@{}rlcccc@{}}
\toprule
agents & search & $f{=}0.25$ & $0.5$ & $0.75$ & mean hop \\
\midrule
50 & directory & .28 & .46 & .76 & 1.0 \\
   & network   & .06 & .26 & .59 & 2.9 \\
250 & directory & .21 & .53 & .76 & 1.0 \\
    & network   & .02 & .10 & .40 & 3.2 \\
1{,}005 & directory & .22 & .42 & .73 & 1.0 \\
        & network   & .01 & .05 & .28 & 3.7 \\
\bottomrule
\end{tabular}
\end{table}

Partial adoption shows the same effect
(Table~\ref{tab:adoption}). With half the agents
using paid mail first, direct search retains 42\%
to 53\% of full adoption, while forwarded search
retains only 26\%, 10\%, and 5\% as networks grow.
Losses increase with the number of inboxes a
request must pass through.

\conclusion{Conclusion 2.4: Inbox failures accumulate across agents}{
A request must be read at each step before it can be
forwarded. When an agent misses a request, other agents
farther along that path cannot be reached. The effects
of individual inbox decisions therefore increase when
coordination requires several agents to communicate.
}

\takeaway{Takeaway 2: The inbox is a coordination primitive}{
Under limited attention and memory, delivering messages
is not enough for agents to coordinate. Each recipient
must decide what to read and retain, and the decisions
affect other agents that depend on its response.
When requests are forwarded, a missed message can also
prevent other agents from being reached. Effective
coordination at scale therefore requires explicit
management of incoming messages, rather than relying
on delivery alone.
}

\subsection{Choosing Inbox Rules}
\label{sec:method:traffic}
\label{sec:eval:policies}

The previous experiments establish the need to manage incoming messages. We now examine how an agent should choose what to read when its attention is limited. A simple rule may be inexpensive but miss important
requests, while using a model to select messages may improve the choice at an additional token cost.

We compare reading rules under different attention
limits and traffic. From 32 incoming messages, an
agent may read 4, 8, or 16 using \emph{newest first},
\emph{paid mail first}, \emph{closest match}, or
\emph{the assistant chooses}. Closest match uses
all-MiniLM-L6-v2 embeddings
\cite{wang2020minilm,reimers2019sbert}, while the
model chooses from one line per message. Controls
either read all messages or only the relevant one.

We test yes or no requests and offers with prices
that must be checked against a private budget.
In the latter setting, half the targets are wanted,
a quarter are similar but unwanted, and a quarter
are unaffordable. Eight competing messages also
carry sender stakes, so paid priority does not
automatically identify the target. We count selection
and judgment costs. Each rule faces 100 separately
drawn cases, so small accuracy differences are not
reliable rankings.

\begin{figure}[t]
\centering
\begin{tikzpicture}
\begin{axis}[paperaxis, width=0.53\columnwidth, height=3.9cm, title={(a) target message read},
  symbolic x coords={4,8,16}, xtick=data, xlabel={messages read, of 32}, ylabel={share of decisions}, ymin=0, ymax=1.05,
  legend to name=legx3, legend columns=2]
\addplot[snewest] coordinates {(4,0.09)(8,0.21)(16,0.53)}; \addlegendentry{newest first}
\addplot[sdir] coordinates {(4,0.39)(8,0.92)(16,1.00)}; \addlegendentry{paid mail first}
\addplot[sclosest] coordinates {(4,0.49)(8,0.53)(16,0.81)}; \addlegendentry{closest match}
\addplot[sassist] coordinates {(4,0.77)(8,0.79)(16,0.96)}; \addlegendentry{the assistant chooses}
\end{axis}
\end{tikzpicture}\hfill
\begin{tikzpicture}
\begin{axis}[paperaxis, width=0.53\columnwidth, height=3.9cm, title={(b) answer right},
  symbolic x coords={4,8,16}, xtick=data, xlabel={messages read, of 32}, ylabel={share of decisions}, ymin=0.4, ymax=1.05]
\addplot[snewest] coordinates {(4,0.52)(8,0.58)(16,0.72)};
\addplot[sdir] coordinates {(4,0.68)(8,0.89)(16,0.98)};
\addplot[sclosest] coordinates {(4,0.80)(8,0.87)(16,0.89)};
\addplot[sassist] coordinates {(4,0.93)(8,0.95)(16,0.95)};
\addplot[sbest, no marks, densely dotted, line width=0.8pt] coordinates {(4,0.93)(16,0.93)};
\node[font=\tiny, color=black!75, anchor=south west, inner sep=1pt] at (axis cs:4,0.93) {best case, read it all (0.93)};
\end{axis}
\end{tikzpicture}\\[2pt]
\pgfplotslegendfromname{legx3}
\caption{No reading rule is best at every budget. 32
arrivals, 100 decisions per cell. (a) Target read. (b) Answer right.
With 4 reads only the assistant choosing finds the target, at nearly
five times the best case's tokens; with 16, paid mail first matches
the best case at a quarter of the assistant's tokens. Gaps under 0.05
are noise.}
\Description{Two line charts against messages read out of 32 (4, 8 and 16) for newest first, paid mail first, closest match and the assistant chooses. (a) Share of decisions in which the target message was read: at four reads only the assistant choosing exceeds one half. At sixteen reads paid mail first reads it every time. (b) Share answered right: at sixteen reads paid mail first matches the best case.}
\label{fig:traffic}
\end{figure}
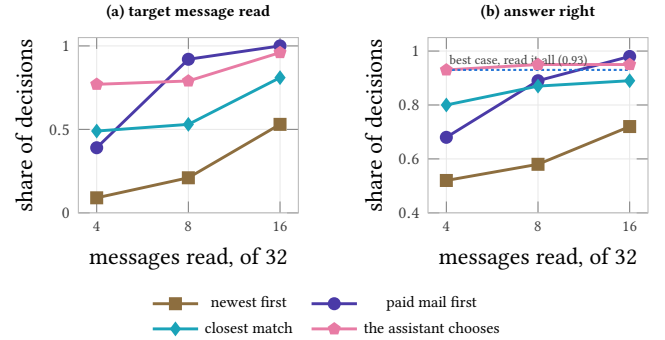

With only four reads, model selection finds the
relevant message in 77\% of cases, against fewer
than half for every other rule
(Figure~\ref{fig:traffic}). However, it costs
nearly five times as many tokens as perfect retrieval.
With sixteen reads, paid mail first finds every
target and matches the accuracy of perfect retrieval
at about a quarter of the tokens of model selection.
At this limit, all staked messages can be read.
With fewer reads, competing staked messages make
priority less reliable.

Private budgets reveal another limitation. Even
model selection reads unaffordable offers as often
as affordable ones, applying the budget only when
responding. A message may therefore appear relevant
without being useful to the recipient.

\takeaway{Takeaway 3: Choosing inbox rules}{
Inbox rules should account for reading limits,
selection costs, and competing traffic. Sender
stakes can provide inexpensive priority when
they distinguish useful requests. Model selection
can help under tighter limits, but costs more
tokens and may overlook private constraints.
}

\subsection{Supporting Runs: Additional Tests of Inbox Rules}
\label{sec:method:supporting}
\label{sec:eval:dilution}
\label{sec:eval:supporting}

We test whether the results persist under partial
adoption, perfect discovery, competing requests,
adversarial stakes, and additional context
(Table~\ref{tab:design}).

\paragraph{Adoption and discovery.}
Under heavy unwanted traffic, a simulated folder
sorter with 89\% classification accuracy reaches
about half the relevant agents, while arrival-order
reading reaches none. The result holds across
adoption rates and two models, with no sudden
change because traffic is added independently
at each agent. Forwarded communication behaves
differently, as shown in
Section~\ref{sec:method:multihop}.
Even a perfect matcher forms no groups under
arrival-order reading in heavy traffic, compared
with about a third under paid mail first.

\paragraph{Competing traffic and stakes.}
Under unwanted traffic, newest first can reach
agents without forming groups. Sorting into
folders also loses performance when competing
messages are genuine requests. Paid priority
depends on sender behavior. An attacker posting
half the required stake gains no priority,
matching the stake reduces the rule to arrival
order, and doubling it prevents genuine requests
from being read.

\paragraph{Additional context.}
With 32 additional messages, response accuracy
falls from 93\% to 77\% for gpt-5.4-mini,
while the larger models show no comparable loss.
In group formation, nearly a third of relevant
agents are consequently missed. Selecting eight
messages recovers much of the loss at less than
half the token cost.

\conclusion{Additional checks: Limits of inbox rules}{
Better discovery cannot ensure that requests
are read. Reading priority depends on competing
traffic and sender behavior, while the effects
of additional context depend on the model.
}

\section{Discussion}

\subsection{Delivery, Discovery, and Reading}

Agent communication involves several distinct problems. Protocols such as MCP and A2A support tool access and agent communication~\cite{anthropic2024mcp,google2025a2a},
while registries such as AGNTCY and NANDA support agent discovery and identity~\cite{muscariello2025agntcy,raskar2025nanda}. Our results concern what happens after a request is delivered. Even a matcher that knows the correct recipients cannot form groups when those recipients fail to read incoming requests under heavy traffic (\autoref{sec:method:supporting}). Finding the correct agent therefore does not guarantee that the request will be considered.

Managing incoming attention is a separate problem from delivery and discovery. Reports of the Hugging Face incident also illustrate the practical risks of large volumes of agent messages~\cite{metr2026hf,openai2026hf}. An inbox determines which delivered messages receive attention and which are left unread.

\subsection{The Inbox is a Coordination Primitive}

Push already has an implicit inbox rule: read messages in arrival order and forget the oldest when memory fills. Our experiments show why the recipient needs greater control. As concurrent interactions increase, reading everything becomes expensive, while forgetting earlier messages can make later decisions inconsistent. The problem depends on the load faced by each agent, rather than society size alone.

The consequences also extend beyond individual agents. When requests are forwarded, each recipient decides whether the request can continue. On a network of 1{,}005 agents with half using paid mail first, direct search retains 42\% of the result under full adoption, while forwarded search retains only 5\% (Table~\ref{tab:adoption}). A rule chosen independently by one agent can therefore affect the ability of other agents to coordinate. We consider the inbox a coordination primitive because receiving and selecting messages are decisions on which other agents depend.

\subsection{Implications for Inbox Design}

\paragraph{Reading.}
The reading rule should depend on available attention and competing traffic. When the reading budget is small, model selection finds more relevant messages but costs more tokens. When the budget is larger, simple priority rules can perform as well at lower cost (\autoref{sec:method:traffic}).

\paragraph{Memory.}
Agents should retain information according to the decisions they need to make. Keeping complete conversations helps an organizer, while an agent managing commitments across conversations benefits from a compact promise list. Even complete history does not guarantee correct decisions, so more memory alone is not always sufficient (\autoref{fig:organizer}, \autoref{fig:shapes}).

\paragraph{Sender stakes.}
A refundable stake can give requests priority without requiring the model to inspect every message. However, a stake indicates willingness to pay, not whether a request is useful. An attacker that matches or exceeds the required stake can defeat a fixed priority rule (\autoref{sec:method:supporting}). Pricing attention
therefore requires considering competing sender behavior.

\paragraph{Replies.}
An agent may answer with a simple decision, a condition, or more detailed private information. The reply vocabulary controls what other agents learn from a response (Table~\ref{tab:inbox}), and hence have privacy implications. Our experiments examine only a limited set of replies, so they do not establish which level of disclosure is best for coordination.

An inbox rule should therefore be evaluated by both the resources it saves for the recipient and the coordination it enables across the society.

\subsection{Limitations and Open Questions}

Our experiments use two simulated coordination tasks
and models from one provider. Identical agent inputs
share model judgments to make large simulations feasible. The results therefore study inbox behavior at large population sizes, rather than independent model decisions by every agent. Experiment~1 was also sensitive to model instructions.

The rules are fixed, and the traffic includes deliberately difficult conditions. In particular, heavy unwanted traffic is chosen to overwhelm arrival-order reading. We do not study agents that adapt their rules over time, or messages designed to manipulate model-based selection~\cite{cohen2025aiworm}. We also evaluate only a small part of the possible reply vocabulary.

Further work should examine adaptive reading and memory rules, sender incentives, and information disclosure across more tasks and heterogeneous agent populations. Our results establish the importance of recipient-side decisions under the tested conditions, rather than the best way to implement them.

\section{Conclusion}
Coordination among language-model agents requires more than delivering messages. As agents manage more concurrent interactions, reading every message becomes expensive, while forgetting information can make later decisions inconsistent. The problem extends beyond individual agents because unread requests can prevent coordination across a network.

We study the inbox as a recipient-controlled layer that decides what to read, what to retain, how senders obtain priority, and what replies may disclose. Our experiments show that these decisions affect coordination, and that their effectiveness depends on agent roles, attention limits, and incoming traffic. The central conclusion is that large agent societies need explicit management of
incoming messages, alongside communication and discovery. The particular inbox rules can evolve with the agents and the systems in which they operate.


\bibliographystyle{ACM-Reference-Format}
\bibliography{references}

\begin{table*}[t]
\caption{Experiment 3. 32 arrivals, of which the agent reads 4, 8 or 16.
100 decisions per cell, gpt-5.6-terra. Yes/no reply: how often the target
was read, how often the answer was right (an unread target counts as
``no''), and tokens per decision (for the assistant, including the call
to choose). Priced reply: how often wanted and too-expensive targets were
read, how often the answer was right, and, for wanted targets read, how
often the stated limit was right.}
\label{tab:traffic}
\small
\setlength{\tabcolsep}{4pt}
\begin{tabular}{@{}lr ccc cccc@{}}
\toprule
& & \multicolumn{3}{c}{yes/no reply} & \multicolumn{4}{c}{priced reply, ``yes, up to \$X''} \\
\cmidrule(lr){3-5}\cmidrule(lr){6-9}
reading rule & reads & target read & right & tokens & read: wanted / too expensive & right & limit right & tokens \\
\midrule
read it all (push) & 32 & 1.00 & 0.93 & 740 & 1.00 / 1.00 & 0.91 & 0.82 & 937 \\
newest first & 4 & 0.09 & 0.52 & 500 & 0.16 / 0.16 & 0.57 & 0.88 & 500 \\
newest first & 8 & 0.21 & 0.58 & 630 & 0.26 / 0.12 & 0.63 & 1.00 & 639 \\
newest first & 16 & 0.53 & 0.72 & 753 & 0.50 / 0.48 & 0.72 & 0.88 & 833 \\
paid mail first & 4 & 0.39 & 0.68 & 434 & 0.44 / 0.36 & 0.68 & 0.82 & 462 \\
paid mail first & 8 & 0.92 & 0.89 & 382 & 0.90 / 0.92 & 0.91 & 0.91 & 485 \\
paid mail first & 16 & 1.00 & \textbf{0.98} & 371 & 1.00 / 1.00 & \textbf{0.94} & 0.88 & 483 \\
closest match & 4 & 0.49 & 0.80 & 410 & 0.68 / 0.68 & 0.83 & 0.97 & 445 \\
closest match & 8 & 0.53 & 0.87 & 521 & 0.74 / 0.68 & 0.86 & 0.97 & 559 \\
closest match & 16 & 0.81 & 0.89 & 598 & 0.82 / 0.84 & 0.89 & 0.95 & 717 \\
the assistant chooses & 4 & 0.77 & \textbf{0.93} & 1{,}177 & 0.96 / 0.96 & \textbf{0.94} & 0.94 & 1{,}364 \\
the assistant chooses & 8 & 0.79 & \textbf{0.95} & 1{,}264 & 0.96 / 1.00 & 0.91 & 0.88 & 1{,}444 \\
the assistant chooses & 16 & 0.96 & \textbf{0.95} & 1{,}341 & 1.00 / 1.00 & \textbf{0.94} & 0.92 & 1{,}589 \\
best case & 1 & 1.00 & 0.93 & 249 & 1.00 / 1.00 & 0.88 & 0.78 & 325 \\
\bottomrule
\end{tabular}
\end{table*}

\appendix
\section{Reproduction and Detail}
\label{app:system}

Every number in the paper regenerates from committed data with one
command. The harnesses, the analysis scripts and every row behind the
tables and figures are released with the paper.

\paragraph{Checks on the model's part.} Every judgment in every
reported run was the model's. In every group-formation run the number
of judgments equals the number of messages delivered minus those
rejected by fixed rules, with no fallback, and every proposal, reply
and choice of what to read was valid output. Agents shown the same
information share one answer, which is why a 16{,}000-agent run makes 6
to 23 model calls, and why one wrong answer applies to every agent in
its group. An earlier calibration on group formation, with the same
protections against leaking the answer, found a carefully worded prompt
on claude-sonnet-5 finding $59\pm4\%$ of the group with 0.1 false
alarms per run, and a naive prompt 42\%. The models used for every
result here find 0.51--0.53 with about two false alarms per run on the
same grids. In meeting scheduling the model writes replies from a
calendar without a malformed answer or a meeting nobody can attend:
yes/no replies match the exact reply on 99.9\% of 6{,}880, coarse on
95.4\% of 4{,}780, full on 97.1\% of 4{,}305. Its mistakes go one way:
it sometimes leaves out a free hour and almost never offers a busy one.

\end{document}